\documentclass[aps,prl,reprint]{revtex4-1}
\usepackage{amsmath}
\usepackage{amssymb}
\usepackage{pxfonts}
\usepackage{graphicx}
\usepackage{eucal}
\usepackage{mathrsfs}
\usepackage{theorem}
\usepackage{pifont}
\usepackage{color}
\usepackage{subfig}
\usepackage{hyperref}

\usepackage{amsmath}
\usepackage{amssymb}
\usepackage{graphicx}
\usepackage{eucal}
\usepackage{mathrsfs}
\usepackage{pifont}
\usepackage{color}
\usepackage{cjhebrew}

\usepackage[all]{xy}
\usepackage{tikz} % Required for drawing custom shapes
\usepackage{tkz-euclide}
\usetkzobj{all}
\usepackage{pgfplots}

\newcommand{\btkz}{\begin{tikzpicture}}
\newcommand{\etkz}{\end{tikzpicture}}

\newcommand{\brk}[1]{\left(#1\right)}          % \brk{.}     => (.)
\newcommand{\Abs}[1]{\left| #1 \right|}        % \Abs{.}     => |.|
\newcommand{\deriv}[2]{\frac{d#1}{d#2}}

\newcommand{\figref}[1]{Figure~\ref{#1}}
\newcommand{\thmref}[1]{Theorem~\ref{#1}}

\newcommand{\beq}{\begin{equation}}
\newcommand{\eeq}{\end{equation}}
\newcommand{\bsplit}{\begin{split}}
\newcommand{\esplit}{\end{split}}
\newcommand{\baligned}{\begin{aligned}}
\newcommand{\ealigned}{\end{aligned}}

\providecommand{\bs}[1]{\boldsymbol{#1}}

\providecommand{\R}{\bbR}

\newcommand{\textand}{\quad\text{ and }\quad}
\newcommand{\Textand}{\qquad\text{ and }\qquad}

\providecommand{\vp}{\varphi}

\newcommand{\Hom}{{\operatorname{Hom}}}

\newcommand{\calA}{{\mathcal A}}

\newcommand{\calM}{{\mathcal M}}

\newcommand{\calQ}{{\mathcal Q}}

\newcommand{\frakg}{\mathfrak{g}}

\newcommand{\frakn}{\mathfrak{n}}

\newcommand{\frakt}{\mathfrak{t}}

\newcommand{\bbR}{{\mathbb R}}

\newtheorem{theorem}{Theorem}

\newcommand{\g}{\frakg}

\newcommand{\Volume}{d\operatorname{Vol}_\g}

\newcommand{\M}{\calM}
\newcommand{\inred}[1]{{\color{red} #1}}
\newcommand{\tvec}{\bs{\frakt}}
\newcommand{\nvec}{\bs{\frakn}}

\newcommand{\x}{\mathbf{x}}

\newcommand{\kernel}{\calQ}
\newcommand{\rhomax}{\rho_{\text{max}}}

\begin{document}

% Use the \preprint command to place your local institutional report
% number in the upper righthand corner of the title page in preprint mode.
% Multiple \preprint commands are allowed.
% Use the 'preprintnumbers' class option to override journal defaults
% to display numbers if necessary
%\preprint{}

%Title of paper
\title{On the Bending Energy of Buckled Edge-Dislocations}

% repeat the \author .. \affiliation  etc. as needed
% \email, \thanks, \homepage, \altaffiliation all apply to the current
% author. Explanatory text should go in the []'s, actual e-mail
% address or url should go in the {}'s for \email and \homepage.
% Please use the appropriate macro foreach each type of information

% \affiliation command applies to all authors since the last
% \affiliation command. The \affiliation command should follow the
% other information
% \affiliation can be followed by \email, \homepage, \thanks as well.
\author{Raz Kupferman}
\email{raz@math.huji.ac.il}
\affiliation{Einstein Institute of Mathematics, The Hebrew University, Jerusalem 91904 Israel}
\thanks{This research was partially funded by the Israel Science Foundation (Grant No. 661/13), and by a grant from the Ministry of Science, Technology and Space, Israel and the Russian Foundation for Basic Research, the Russian Federation.}

%Collaboration name if desired (requires use of superscriptaddress
%option in \documentclass). \noaffiliation is required (may also be
%used with the \author command).
%\collaboration can be followed by \email, \homepage, \thanks as well.
%\collaboration{}
%\noaffiliation

\date{\today}

\begin{abstract}
The study of elastic membranes carrying topological defects has a longstanding history, going back at least to the 1950s. When allowed to buckle in three-dimensional space, membranes with defects can totally relieve their in-plane strain, remaining with a bending energy, whose rigidity modulus is small compared to the stretching modulus.
In this paper, we study membranes with a single edge-dislocation. We prove that the minimum bending energy associated with strain-free configurations diverges logarithmically with the size of the system.
\end{abstract}

% insert suggested PACS numbers in braces on next line
\pacs{\inred{TBD}}
% insert suggested keywords - APS authors don't need to do this

%\maketitle must follow title, authors, abstract, \pacs, and \keywords
\maketitle

%%%%%%%%%%%%%%%%%%%%%%%%%%%%%%%%%%%%%%%%%%
\section{Introduction}

The energetics of two-dimensional (2D) elastics membranes with defects has been studied extensively in the past several decades (e.g., \cite{ES51,NP87}). In a crystalline setting, one may model a 2D solid with a single disclination by a triangular lattice perturbed by a unique vertex of degree either $5$ (positive disclinations) or $7$ (negative disclinations) \cite{SN88}. Likewise, a 2D solid with a single dislocation can be modeled by a triangular lattice perturbed by a five-seven pair. In a continuum setting, the geometry of the elastic membrane is modeled by a Riemannian metric describing local equilibrium distances between neighboring material elements. The ordered state is modeled by a Euclidean metric, implying that the membrane can be embedded locally in Euclidean plane without stretching. Defects are modeled by singularities in the metric \cite{Vol07}: a disclination corresponds to a Dirac measure-valued Gaussian curvature, whereas a dislocation corresponds to a dipole of Gaussian curvature. The presence of defects constitutes a \emph{metric incompatibility} between the intrinsic geometry of the membrane and planar geometry.

When confined to planar configurations, an elastic membrane with defects of either type will necessarily be metrically-distorted. In a crystalline setting, distortions manifest as a stretching or a compression of lattice bonds;   in a continuum setting, distortions manifest as deviations of the \emph{actual metric} of the membrane from its \emph{reference metric}. The elastic energy of a configuration is a measure of this metric distortion. It is evidently model-dependent, however, prototypical models assume an elastic energy that scales quadratically with the local distortion. 

In this paper, we focus on membranes with single dislocations. It is well-known that the elastic (stretching) energy $E_S$ of planar configurations is bounded from below by a term depending quadratically on the magnitude of the dislocation (the Burgers vector), and diverging logarithmically both with the linear size of the system, $R$, and the radius $r_0$ of a core region around the defect, which can either be removed from the model, or regularized; this scaling is sharp, in the sense that an upper bound with the same scaling can be obtained for low-energy configurations. 

If allowed, thin membranes can buckle in the three-dimensional (3D) ambient space.\cite{ES51,NP87,SN88}. Buckling allows for the full relaxation of the stretching energy $E_S$; in geometric terms, this means that up to some finite core, a surface with a dislocation can be embedded in 3D Euclidean space isometrically. From an energetic point, stretching energy $E_S$ is being traded for a \emph{bending energy} $E_B$, which is a higher-order measure of distortion, where the relevant small parameter if typically the ratio of the membrane thickness $t$ and the system size $L$; in certain cases, e.g., in graphene, an effective measure of plate thickness is defined by the ratio of bending and stretching moduli.
%While the in-plane stretching energy of buckled configuration may be zero, 
%a higher-order energy contribution is typically assumed in the form a \emph{bending energy} $E_B$, whose rigidity modulus scales, in a finite-thickness setting with the second power the membrane thickness, $t$. 
The bending energy is related to the so-called Willmore functional, i.e., the surface integral of the membrane's mean curvature squared. There exists a vast literature on the dimensional reduction of 3D elasticity into so-called plate, shell and membrane models, starting from phenomenological arguments \cite{Kar10}, through asymptotic analyses \cite{Lov27,Koi66}, and more recently, rigorous limit theorems \cite{LR95,FJM02b}; in the metrically-incompatible context, an asymptotically-based argument was presented in \cite{ESK09a}, followed by rigorous analyses in \cite{LP10,KS14}. 

A question of both fundamental and practical importance (e.g., the melting transition in 2D membranes \cite{NP87,Nel02}) is whether low-energy configurations of buckled dislocations remain finite as $R$ tends to infinity. A natural reference system is that of a disclination, which may also be embedded isometrically in 3D Euclidean space, however, with a bending energy diverging logarithmically with the $R$; see \cite{Olb17,Olb18} for recent rigorous analyses departing from 3D models. Heuristic arguments have suggested that in dislocations, which are bound pairs of disclinations of opposite signs, the logarithmic contributions may cancel out giving rise to an energy bound independent of $R$. Numerical simulations were performed in \cite{SN88}, supporting these heuristics.  

It should be noted that there is a certain degree of fuzziness in the statement that low-energy configurations are independent of $R$. Starting either from a full 3D model, or from a Koiter plate model \cite{Koi66}, 
\[
E_{\text{total}} = E_S + t^2 E_B,
\]
two distinct limits may be considered: the plate limit $t\to0$ and the ``thermodynamics" limit $R\to\infty$. It is not at all clear that these two limits are interchangeable. If $t$ is taken to zero first for finite $R$ and a removed core, theorems establish that the low-energy states are isometric immersions, i.e., states of zero stretching energy. After rescaling the energy by $t^2$, the leading order energy is the bending energy, now restricted to the space of isometric immersions. A key question is whether the minimal bending energy remains finite as $R\to\infty$. Alternatively, one may let $R\to\infty$ first for finite $t$. Assuming that finite-energy states exist (possibly combining both stretching and bending contributions), one could study the $t\to0$ limit. A third alternative would be to let $t\to0$ and $R\to\infty$ simultaneously, assuming a certain relation between both variables. 

It is not clear to what extent these distinct alternatives have been recognized in the literature. Most references mention the fact that out-of-plane buckling allows for the complete elimination of stretching energy. It seems a common belief that the bending energy of strain-free buckled dislocations is either $R$-independent, or, to the least, diverges with $R$ slower than logarithmically.

We prove that this is not the case; we show that the bending energy can be bounded from below by a term diverging logarithmically with $R$, being in this sense, similar to a disclination (even though, the two cases differ substantially, as will be discussed). Specifically, our main result is the following:

\begin{theorem}
\label{th:main}
Consider a 2D annulus of inner radius $r_0$ and outer radius $R$, endowed with a metric representing an edge-dislocation with Burgers vector $b$. Then, the bending energy of isometric immersions of that surface into $\R^3$ is bounded from below by
\beq
E_B \ge \frac{b^2}{128\pi^3 r_0^2} \log \frac{R}{r_0}.
\label{eq:final_bound}
\eeq
\end{theorem}

%%%%%%%%%%%%%%%%%%%%%%%%%%%%%%%%%%%%%%%%%%
\section{Geometry of an edge-dislocation}

The geometry of a single edge-dislocation can be defined independently of any parametrization \cite{KMS15,MLAKS15}. The membrane is modeled as a 2D Riemannian manifold $(\M,\g)$ having an annular topology. The metric $\g$ is locally-Euclidean---every point $p\in\M$ has an open neighborhood isometrically embeddable in Euclidean plane. A locally-Euclidean geometry implies a flat (Levi-Civita) connection $\nabla^\M$, or equivalently, a locally path-independent parallel transport. The difference between a disclination and a dislocation is that in the latter case, the net curvature ``inside the whole" is zero (trivial holonomy), namely, parallel transport is globally path-independent. We denote the parallel transport by $\Pi_p^q:T_p\M\to T_q\M$ for $p,q,\in \M$. The presence of a dislocation is additionally reflected by a non-zero circulation: there exists a $\nabla^\M$-parallel vector field $b\in\Gamma(T\M)$, such that for every closed loop $\gamma:I\to\M$ encircling the core (homotopic to the inner-boundary) and 
for every reference point $p\in\M$,
\[
\int_I \Pi_{\gamma(t)}^p(\dot{\gamma}(t))\, dt = b_p,
\]
where the evaluation of a vector field at a point is denoted by a subscript, as in $b_p$ (this integral is the continuum counterpart of the lattice step counting in crystalline solids). Finally, the size of the system is imposed by setting the geodesic curvatures of the inner and outer boundaries to be close to $r_0$ and $R$ respectively. In edge-dislocations, as opposed to screw-dislocations, the magnitude of the Burgers vector cannot exceed the perimeter $2\pi r_0$ of the inner boundary.

While the geometry of an edge-dislocation is well defined (up to immaterial details) by the above characterization, a coordinate representation is often more suitable for calculations. A convenient coordinate representation is the following: use polar-like coordinates,
\[
(r,\vp) \in [r_0,R]\times[0,2\pi),
\]
where periodicity in $\vp$ is assumed. 
In these coordinates, the metric $\g$ is given by
\[
\g(r,\vp) = dr\otimes dr +  (r_0 + (r-r_0)\kappa)^2 d\vp\otimes d\vp,
\]
where $\kappa(\vp) = 1 + B\,\cos\vp$ and $|B|<1/2$ is a dimensionless parameter related to the ratio of the Burgers vector and the core size. We note that the frame field $\{e_1,e_2\}$, with
\[
e_1 = \partial_r
\Textand
e_2 = [r_0 + (r-r_0)\kappa]^{-1}\partial_\vp,
\]
is orthonormal.
The geodesic curvature of constant-$r$ curves  is
$k_g(\vp) =  \kappa(\vp)/r$, and their perimeter is $2\pi r$.
One may furthermore verify that the Gaussian curvature vanishes locally, i.e., the manifold is locally-Euclidean, and that the holonomy is trivial. Finally, the Burgers vector equals $b = -2\pi r_0 J_1(B)\, e_2$, where $J_1$ is the Bessel function of the first kind (see Supp. Mat. for details). 

\begin{figure}
\includegraphics[height=1.6in]{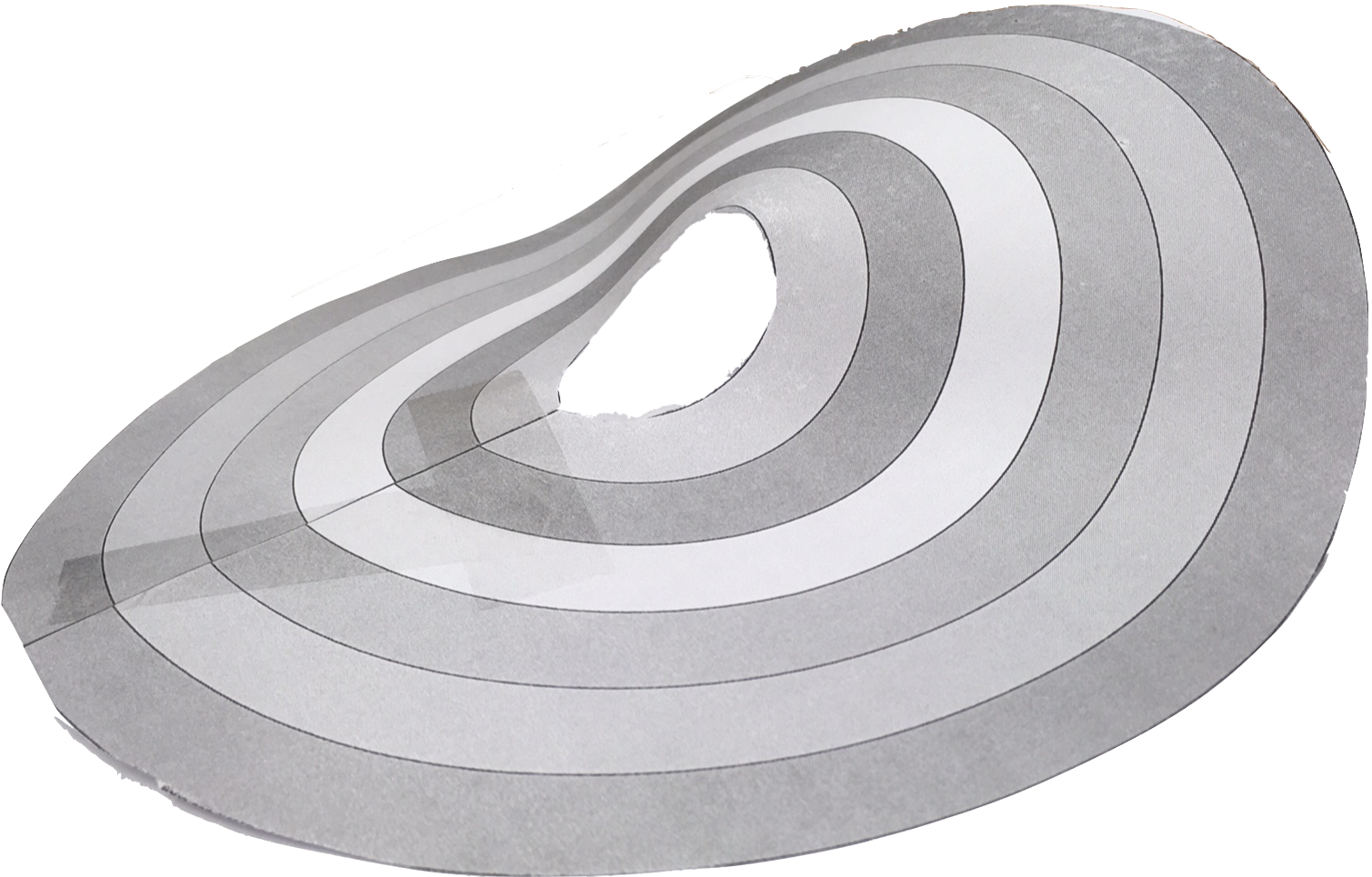}
\caption{Image of a buckled membrane with an edge-dislocation.}
\label{fig:0}
\end{figure}

A configuration of $(\M,\g)$ is an immersion $\x:\M\to\R^3$, where $\R^3$ is endowed with the standard Euclidean metric; an image of a configuration of an edge-dislocated paper sheet is shown in \figref{fig:0}.
Denoting by $H:\M\to\R$ the mean curvature of $\x(\M)$ in $\R^3$, the bending energy associated with $\x$ is given by
\beq
E_B(\x) = \int_\M H^2\,\Volume,
\label{eq:functional}
\eeq
where $\Volume$ is the area element induced by $\g$. 
Here we focus on membranes with effectively zero thickness, where only bending deformations are allowed. Therefore, our goal is to find a lower bound for $E_B$ over all isometric immersions $\x:\M\to\R^3$.

%%%%%%%%%%%%%%%%%%%%%%%%%%%%%%%%%%%%%%%%%%
\section{Proof of \thmref{th:main}}

In this section we derive the lower bound \eqref{eq:final_bound}. First, there is an analytical subtlety that needs to be addressed: the functional \eqref{eq:functional} is naturally defined on the space of Sobolev functions $W^{2,2}(\M;\R^3)$; correspondingly, derivatives should be interpreted in a weak sense.  Hornung \cite{Hor08} (building upon Pakzad \cite{Pak04}) proved that the set of smooth isometric immersions is dense in the $W^{2,2}$ topology within the set of $W^{2,2}$ isometric immersions. Thus, the infimum of $E_B$ over smooth isometric immersions is the same as its infimum over $W^{2,2}$ isometric immersions. In other words, the bending energy for isometric immersions cannot be lowered by deteriorating the regularity. In practical terms, this implies that we may restrict our analysis to smooth maps. 

Fix $r\in[r_0,R]$, and consider the constant-$r$ curve in arclength parametrization
$\gamma:[0,2\pi r)\to \M$;  clearly, $\dot{\gamma} = e_2$. 
By the definition of the Burgers vector
\beq
\int_0^{2\pi r}  \Pi_{\gamma(t)}^{\gamma(0)}(e_2)\, dt = b_{\gamma(0)}.
\label{eq:surf7}
\eeq
On the other hand, denote by $\varpi^{y}_x:T_x\R^3\to T_{y}\R^3$ the parallel transport in $\R^3$; since $\R^3$ is defect-free,
\beq
\int_0^{2\pi r}  \varpi_{\x(\gamma(t))}^{\x(\gamma(0))} (d\x_{\gamma(t)}(e_2))\, dt = 0.
\label{eq:surf8}
\eeq
Operating with $d\x_{\gamma(0)}$ on \eqref{eq:surf7} and subtracting \eqref{eq:surf8}, we obtain
\beq
\int_0^{2\pi r}  \kernel(t)(e_2)\, dt = d\x_{\gamma(0)}(b_{\gamma(0)}),
\label{eq:surf11}
\eeq
where 
\[
\kernel(t) = d\x_{\gamma(0)}\circ \Pi_{\gamma(t)}^{\gamma(0)} -
\varpi_{\x(\gamma(t))}^{\x(\gamma(0))} \circ d\x_{\gamma(t)}.
\]
Taking (Euclidean) norms in \eqref{eq:surf11}, and using the fact that $d\x$ is an isometry, 
\beq
|b|\le  \int_0^{2\pi r}  \Abs{\kernel(t)}\, dt,
\label{eq:surf9}
\eeq
where the norm of $K(t)\in \Hom(T_{\gamma(t)}\M, T_{\x(\gamma(0))}\R^3)$ is induced by $\g$ and the Euclidean metric;
note that we write $|b|$ rather than $|b_{\gamma(0)}|$, since $b$ is a parallel field, hence its norm is the same everywhere.

We proceed to estimate the integrand on the right-hand side of \eqref{eq:surf9}. Differentiating $|\kernel(t)|^2$ with respect to $t$, 
\[
\begin{split}
\deriv{|\kernel|^2}{t} &= 
2\sum_{i=1}^2 \brk{\kernel(e_i),
d\x_{\gamma(0)}\circ \Pi_{\gamma(t)}^{\gamma(0)} \circ \nabla^\M_{e_2} (e_i)} \\
&- 2\sum_{i=1}^2 \brk{\kernel(e_i),
\varpi_{\x(\gamma(t))}^{\x(\gamma(0))} \circ \x^*\nabla^{\R^3}_{e_2} \circ d\x_{\gamma(t)}(e_i) },
\end{split}
\]
where $\x^*\nabla^{\R^3}$ is the pullback of the Euclidean connection. Substituting the definition of the second fundamental form of $\x(\M)$ in $\R^3$,
\[
\operatorname{II}(u,v) = (\x^*\nabla^{\R^3})_{u} d\x(v) - d\x(\nabla^{\M}_u v), \quad u,v\in T\M,
\]
and the expression for the Cartan-Christoffel symbols (Supp. Mat., Eq. (SM2)), we obtain after straightforward manipulations,
\[
\deriv{|\kernel|^2}{t} = 
2\sum_{i=1}^2\brk{\kernel(e_i),
\varpi_{\x(\gamma(t))}^{\x(\gamma(0))} \circ \operatorname{II}(e_2,e_i)}.
\]
Using the Cauchy-Schwarz inequality and the fact that $\varpi_x^y$ is an isometry, 
\beq
\deriv{}{t} |\kernel| \le |\operatorname{II}|.
\label{eq:surf17}
\eeq
Since the surface is locally-Euclidean, the norm of the second fundamental form coincides with the absolute mean curvature, $|H|$. Furthermore, since $\kernel(0)=0$, it follows from \eqref{eq:surf17} that
\[
|\kernel(t)| \le \int_0^t |H(\gamma(s))|\, ds \le  \int_0^{2\pi r} |H(\gamma(s))|\, ds.
\]
Substituting into \eqref{eq:surf9},
\beq
\frac{|b|}{2\pi r} \le  \int_0^{2\pi r} |H(\gamma(s))|\, ds.
\label{eq:surf14}
\eeq
Squaring and applying once again the Cauchy-Schwarz inequality, we finally obtain
\beq
\frac{|b|^2}{8\pi^3 r^3} \le  \int_0^{2\pi r} H^2(\gamma(s))\, ds.
\label{eq:surf12}
\eeq

Equation \eqref{eq:surf12} is a lower bound on the integral of the mean curvature square along a constant-$r$ loop. Since the left-hand side decays like $1/r^3$, integration over $r$ yields a lower bound for $E_B$ independent of $R$; this situation is very different than in disclinations, where a similar analysis yields a left-hand  side proportional to $1/r$ (see Supp. Mat., Eq. (SM12)), hence a bending energy with lower bound diverging logarithmically with $R$. Note that the difference between the two cases could have been anticipated by a simple dimensional argument. 

Nevertheless, it would be premature to infer that the bending energy can be bounded independently of $R$. We have only learned that a diverging lower bound cannot be obtained by segmenting the annulus into annular stripes and summing up  energy bounds for each stripe.

We proceed to the second part of the analysis, which consists of relating the bending content of the inner boundary with the bending content of loops inside the body. Denote by $\x_0(s)\in \R^3$, $s\in[0,2\pi r_0)$ the configuration of the inner boundary in arclength parametrization; denote by $\tvec(s) = \x_0'(s)$ is unit tangent. As is well-known (e.g., \cite{Mas62}), a locally-flat surface in $\R^3$, is a \emph{developable surface}: It can be partitioned into flat points (where $H=0$) and non-flat points, the latter constituting an open set. Through every non-flat point passes a unique asymptotic line---a geodesic (in $\M$) which maps under $\x$ into a geodesic (in $\R^3$). Asymptotic lines do not intersect, and do not terminate until hitting the boundary.

\begin{figure}
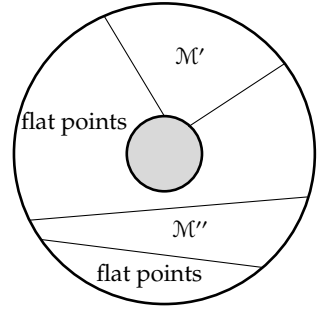

\btkz
	\draw[line width=1pt] (0,0) circle (2cm);
	\draw[line width=1pt,fill=gray!30] (0,0) circle (0.5cm);
	\draw (0.32,0.36) -- (1.6,1.2);
	\draw (0.,0.5) -- (-0.8,1.83);
	\node at (0.35,1.3) {$\M'$};
	\draw (-1.80,-0.885) -- (1.90,-0.575);
	\draw (-1.65,-1.14) -- (1.32,-1.515);
	\node at (0.35,-1.) {$\M''$};
	\node at (-0.2,-1.62) {{\small flat points}};
	\node at (-1.2,0.4) {{\small flat points}};
\etkz
\caption{Partition of $\M$ into flat and non-flat points. $\M'$ is the set of non-flat points connected by asymptotic lines to the inner boundary. $\M''$ is the set of non-flat points connected  by asymptotic lines only to the outer boundary.}
\label{fig:1}
\end{figure}

The immersion $\x$ induces a semi-geodesic parametrization of an open submanifold $\M'$ of $\M$. Specifically, 
let $\calA\subset[0,2\pi)$ be the set of values $s$ for which $\x_0(s)$ is a non-flat point; 
set $\M'$ to be the union of the asymptotic lines emanating from $\calA$; see \figref{fig:1}.
We parametrize $\M'$ with $s\in\calA$ and with the arclength $\rho$ along the asymptotic line; $\rho$ ranges from $0$ to some $\rhomax(s)$ ranging between $R$ and $R+r_0$, and depending on the angle between $\partial_s$ and $\partial_\rho$. 
For every $s\in\calA$, let 
$\nvec(s)\in\R^3$ be unit vector along the embedded asymptotic line emanating through $\x_0(s)$. 
The restriction of $\x$ to $\M'$ is given, by construction, by
\beq
\x(\rho,s) = \x_0(s) + \rho\, \nvec(s).
\label{eq:form_of_f}
\eeq
%It should be emphasized that $\M'$ consists of all those non-flat points in $\M$, which are connected to the inner boundary by an asymptotic line; see \figref{fig:1}.

We proceed to claim that with no loss of generality, we may assume that $\M'$ contains \emph{all} the non-flat points in $\M$. Indeed, consider, for example, the region marked $\M''$ in \figref{fig:1}, and assume it is a connected component of the set of non-flat points; as proved in \cite{Mas62}, such a set, along with its boundary, is a union of $\M$-geodesic, mapped by $\x$ into straight lines. This region can be flattened without affecting the mean curvature in any other region, i.e., the bending energy can be reduced, without changing the restriction of $\x$ to $\M'$.

The metric induced by an immersion of the form \eqref{eq:form_of_f} has entries
\[
E = 1
\qquad
F = \tvec \cdot \nvec
\textand
G = 1 + 2\rho  \, \tvec \cdot \nvec' + \rho^2 |\nvec'|^2,
\]
where we used the fact that $\nvec\cdot\nvec'=0$ (see, e.g., \cite{Str61} for a standard notation of the first and second fundamental forms). The unit vector $\nvec$ cannot be chosen independently of $\x_0$; it follows from the Brioschi formula that the Gaussian curvature vanishes if and only if $\tvec$, $\nvec$ and $\nvec'$ are coplanar (see Supp. Mat.). 

The second fundamental form of $\x(\M')$ in $\R^3$ is not an isometric invariant. By construction, the entries $e$ and $f$ of the second fundamental form vanish. The third entry $g$ can be expressed as a function of $\rho$ and the functions $\tvec\cdot\nvec$ and $\tvec'\cdot\nvec$ of $s$; it follows directly from the Codazzi-Mainardi compatibility conditions \cite[p. 111]{Str61} that  $g/\sqrt{EG-F^2}$
is independent of $\rho$, i.e., it is constant along asymptotic lines; expressing the mean curvature $H$ in terms of the two fundamental forms, we obtain (see Supp. Mat.) that
\beq
H\,\sqrt{EG-F^2} \quad\text{is independent of $\rho$}.
\label{eq:H_indep_rho}
\eeq

It remains to combine \eqref{eq:H_indep_rho} together with the lower bound \eqref{eq:surf14} to obtain a lower bound for the total bending energy. As shown in the Supp. Mat., a combination of the two yields the lower bound \eqref{eq:final_bound}.

%%%%%%%%%%%%%%%%%%%%%%%%%%%%%%%%%%%%%%%%%%
\section{Discussion}

We proved that the minimal bending energy of a strain-free buckled dislocation diverges logarithmically with the size of the system. This result may be surprising for two reasons: (i) the conjecture, whereby the logarithmic divergence associated with two disclinations of opposite signs may cancel out, turns out to be incorrect;  (ii) the scaling of the energy bound is the same as for disclinations. Note, however, the substantial difference between the two cases: For disclinations, the energetic contribution of a stripe at a distance $r$ from the core scales like $O(1/r)$, whence the logarithmic divergence. For dislocations, the energetic contribution of a stripe scales like $O(1/r^3)$; the logarithmic divergence results from the propagation of curvature within the manifold. While the main focus here has been on the $R$-dependence of the bending energy, note the substantial difference in the dependence on $r_0$: for disclinations, the bending energy of an isometric immersion is bounded from below by
\[
C\,\log\frac{R}{r_0},
\]
where the dimensionless prefactor $C$ depends on the magnitude of the disclination. Thus, increasing $r_0$ while retaining the defect intensity fixed has a mild effect in the case of a disclination, whereas for dislocations, the bending energy decreases with $r_0$ quadratically.  
The distinction between disclinations and dislocations has a practical implication: if a cone is segmented into a set of narrow circular conical stripes, the total energy of all the segments when separated from each other is equal to the energy of the cone as a whole. This is not the case for a dislocation, where segmentation results in energetic relaxation.

Another interesting observation is the different scalings of bending and stretching energies in dislocations, assuming that the core radius $r_0$ and the Burgers vector $b$ are of the same order. Then,
\[
\begin{gathered}
E_S=0 \quad\Rightarrow\quad t^2 E_B \sim t^2 \log\frac{R}{b} \\
E_B=0 \quad\Rightarrow\quad E_S \sim b^2 \log\frac{R}{b}.
\end{gathered}
\]
As to be expected, buckling is preferable only as long as the body is thin, i.e., $t$ is smaller than all other intrinsic lengths.

As exposed in the Introduction, the order in which the limits $h\to0$ and $R\to\infty$ are taken is substantial. 
The case where $t\to0$ first and then $R\to\infty$ is well-understood.  A limiting behaviour, which to the best of my knowledge is not yet understood, is the case of finite thickness and infinite radius, letting then $t\to0$. For such a  case to make sense, one would first need to show that there exist configurations for which combined stretching and bending remain finite as $R\to\infty$. While the existence of such configurations is not doubted in the physics literature, a rigorous existence proof is still lacking.

\paragraph{Acknowledgments}
I am indebted to Michael Moshe for introducing me to this problem and for his invaluable advice.
I have benefitted from discussions with Cy Maor and from his critical reading of the manuscript.

%%%%%%%%%%%%%%%%%%%%%%%%%%%%%%%%%%%%%%%%%%
%\bibliography{/Users/raz/Dropbox/tex/Refs/MyBibs}

\end{document}